\documentclass[aps,amsmath,amssymb,twocolumn,longbibliography,showpacs,floatfix]{revtex4-1}

\usepackage{graphicx}
\usepackage{dcolumn}
\usepackage{bm}
\usepackage{color}
\usepackage{ulem}
\usepackage[colorlinks=true,allcolors=blue]{hyperref}%

\begin{document}

\title{Influence of electron-vibration interactions on electronic current noise \\ of atomic and molecular junctions}

\author{S. G. Bahoosh$^{1}$}
\author{M. A. Karimi$^{1}$}
\author{W. Belzig$^{1}$}
\author{E. Scheer$^{1}$}
\author{F. Pauly$^{2,1}$}
\email{Corresponding author: fabian.pauly@oist.jp}

\affiliation{$^{1}$Department of Physics, University of Konstanz, 78457 Konstanz, Germany}
\affiliation{$^{2}$Okinawa Institute of Science and Technology Graduate University, Onna-son,
Okinawa 904-0495, Japan}

\date{\today}

\begin{abstract}
We present an ab-initio method to simulate the current noise in the presence of electron-vibration interactions in atomic and molecular junctions at finite temperature. Using a combination of nonequilibrium Keldysh Green's function techniques and density functional theory, we study the elastic and inelastic contributions to electron current and shot noise within a wide range of transmission values in systems exhibiting multiple electronic levels and vibrational modes. Within our model we find the upper threshold, at which the inelastic noise contribution changes sign, at a total transmission between $\tau\approx 0.90$ and $0.95$ for gold contacts. This is higher than predicted by the single-level Holstein model but in agreement with earlier experimental observations. We support our theoretical studies by noise measurements on single-atom gold contacts which confirm previous experiments but make use of a new setup with strongly reduced complexity of electronic circuitry. Furthermore, we identify 1,4-benzenedithiol connected to gold electrodes as a system to observe the lower sign change, which we predict at around $\tau\approx 0.2$. Finally, we discuss the influence of vibrational heating on the current noise.
\end{abstract}

\maketitle

\section{Introduction}

Shot noise is a nonequilibrium fluctuation phenomenon, which is caused by the discreteness of charge carriers \cite{Blanter2000,Beenakker2003,Cuevas2017}. When the size of an electronic system reaches the nanometer scale, it becomes an important tool for exploring correlations in charge transport \cite{DePicciotto1997,Saminadayar1997,Cron2001,Zhou2019} or for detecting temperature differences on the atomic scale \cite{Lumbroso2018}. Shot noise reveals additional information on the electronic structure and transport properties, like the number of open transmission eigenchannels and their transmissions, which cannot be obtained by conventional studies of the conductance \cite{Reznikov1995,Brom1999,Djukic2006,Kiguchi2008,Schneider:2010ek,Schneider:2012cb,Kumar2013,Ben-Zvi2013,Vardimon2013,Karimi2016a}. Besides shot noise, also low-frequency flicker noise contains valuable information on the type of transport in nanoscale systems, e.g.\ on ballistic versus diffusive transport or two-level current fluctuations \cite{Santa2019,Xiang2015}. When the applied voltage vanishes, the Johnson-Nyquist thermal noise $4k_{\textrm B}TG$ dominates, where $k_{\textrm B}$ is the Boltzmann constant, $T$ the temperature and $G$ the linear conductance. Once the system is biased, it is gradually moved away from equilibrium, and a crossover from thermal to nonequilibrium noise occurs. Assuming phase-coherent elastic charge transport, as described in the framework of Landauer-B\"uttiker scattering theory, the conductance can be expressed as
\begin{equation}
    \label{eq:G}
    G=G_0\tau
\end{equation}
 and the noise power as \cite{Buettiker1992,Blanter2000,Cuevas2017}
\begin{align}
  \begin{aligned}
    S(V)  = & 2eVG_0 \coth\left(\frac{\beta eV}{2}\right)\sum_{n=1}^N
    \tau_n(1-\tau_n)\\
    &+4k_{\text{B}}TG_0\sum_{n=1}^N \tau_n^2,
  \end{aligned}
  \label{eq:Sel}
\end{align}
where $e=|e|$ is the elementary charge, $G_0=2e^2/h$ the quantum of conductance, $V$ the applied bias voltage, $0\leq \tau_n\leq 1$ the transmission probability of the $n$th transmission eigenchannel, $N$ the number of allowed transmission eigenchannels, and $\beta=1/(k_{\text{B}}T)$. In Eq.~\eqref{eq:G} we use the total transmission 
\begin{equation}
    \tau=\sum_{n=1}^N\tau_n.
\end{equation}
When $k_{\text{B}}T\ll eV$, the noise power is reduced to its classical shot noise form $S=2e F I$. Here $I$ is the time-averaged current, while 
\begin{equation}
  F=\frac{\sum_{n=1}^N \tau_n(1-\tau_n)}{\sum_{n=1}^N \tau_n}
  \label{eq:Fano}
\end{equation}
is the Fano factor that describes the dependence of $S$ on the individual transmission probabilities $\tau_n$.

From Eq.~\eqref{eq:Sel} it is clear that for a perfect transmission of all channels ($\tau_n=1$ for $n=1,\ldots,N$) shot noise $S(V)-S(V=0)$ is suppressed, because there are no fluctuations in the occupation numbers of left- and right-moving electrons. This picture of elastic transport with non-interacting electrons was confirmed experimentally for nanostructures and mesoscopic conductors in general on many occasions \cite{Blanter2000,Agrait2003,Cuevas2017,Song2017,Evers2019}.

Nontrivial deviations from elastic scattering theory are expected as the bias voltage exceeds the threshold for excitation of vibrational modes. The study of electron-vibration (EV) scattering in nanoscale systems, especially atomic and molecular junctions, has become an important tool to determine the precise contact geometry and to identify the molecule that bridges two metal electrodes. It can be addressed experimentally by point-contact spectroscopy \cite{Agrait2002,Smit2002,Agrait2003}, inelastic electron tunneling (IET) spectroscopy \cite{Kushmerick2004,Wang2004,Galperin2004,Smit2002,Djukic2006,Song2009,Kim2011a,Karimi2016b} or Raman spectroscopy \cite{Ioffe2008,Ward2010,Bi2018}. Point-contact and IET spectroscopy are identical techniques, but the names originate from the different high and low conductance ranges, respectively, in which they are applied.

Theory has made major contributions to the better understanding of the influence of inelastic EV interactions on charge current. Using a Hamiltonian with a single electronic level coupled to two electrodes and a single vibrational mode, which we will refer to in the following as single-level Holstein model (SLHM), observations of point-contact and IET spectroscopy could be understood in a unified framework \cite{Vega2006,Paulsson2005,Paulsson2008,Cuevas2017}. Thus, a transition from a step up to a step down in the differential conductance has been predicted with increasing transmission as the bias voltage is swept across the vibrational energy from below. For symmetric single-channel junctions and weak EV coupling the sign change in the differential conductance takes place at $\tau=1/2$, with a corresponding decrease for asymmetrically coupled junctions \cite{Vega2006,Paulsson2005,Paulsson2008,Cuevas2017}. Subsequent theory work was directed towards inclusion of strong EV couplings \cite{Schinabeck2016} and towards material-specific predictions that take multiple electronic levels and vibrations as well as the coupling between them into account \cite{Frederiksen2007,Buerkle2013}. To avoid free parameters, ab-initio electronic structure theory, in particular DFT, has been employed to obtain the electronic Hamiltonian, vibrational modes and corresponding EV couplings \cite{Frederiksen2007,Buerkle2013}. The theoretical predictions of the sign change in inelastic conductance corrections have been confirmed experimentally for different systems \cite{Tal2008,Kim2011a}.

Similar to point-contact, IET or Raman spectroscopy of the current, inelastic current noise of atomic-scale junctions provides unique information on the system such as the local phonon population. Therefore the studies of inelastic EV interactions have subsequently been extended to shot noise. Using the SLHM and assuming weak EV coupling, two transitions at different transmissions have been predicted, at which the step in the first derivative of the inelastic noise contribution with respect to voltage changes sign at the vibrational energy \cite{Avriller2009,Schmidt2009,Haupt2009}. In the case of a symmetrically coupled single-channel junction with $\tau=\tau_1$ the resulting step, as the bias crosses the vibrational energy, is positive, when $\tau < \tau_-=(1-1/\sqrt{2})/2 \approx 0.15$ or $\tau > \tau_+=(1+1/\sqrt{2})/2)\approx 0.85$, and negative in between \cite{Avriller2009}. The predictions remain basically unchanged, if a scattering-state formulation is adopted \cite{Kim2014}. For a more realistic picture, the scheme of Ref.~\cite{Haupt2009} has later been extended to be compatible with the existing ab-initio approaches, applied to point-contact and IET spectra, as described above. Using the lowest-order expansion of the shot noise in the EV coupling, it takes multiple electronic levels, vibrational modes and all relevant EV couplings into account \cite{Haupt2010}. This formulation was utilized in Ref.~\cite{Avriller2012} to study both the elastic and inelastic shot noise in gold (Au) and platinum (Pt) metallic atomic contacts within a parameter-free DFT simulation. Inelastic processes strongly alter the high-voltage behavior \cite{Novotny2011,Novotny:2015jy}. 

While shot noise in atomic contacts and molecular junctions has already been
measured in the elastic low-bias regime
\cite{Brom1999,Smit2002,Djukic2006,Kiguchi2008,Vardimon2013,Karimi2016a}, the
inelastic effects on electronic shot noise due to EV couplings have been
analyzed only recently
\cite{Tsutsui2010,Kumar2012,Chen2014,Xiang2015,Chen2016,Tewari2018,Tewari2019}. Tsutsui
et al.~\cite{Tsutsui2010} investigated the fluctuations of charge current
flowing through a single molecule at $T=4$~K. They report an increased noise
signal at voltages that are characteristic for the molecular vibrational modes
and find a similar peak structure in both shot noise and IET spectra. Tewari
et al.~\cite{Tewari2019} recently demonstrated for deuterium molecules that
the inelastic noise contributions due to vibrational excitations interacting
with two-level fluctuators can be used to detect inelastic processes more
precisely than with IET spectra.  In the experimental work by Kumar et
al.~\cite{Kumar2012} the shot noise of Au atomic contacts was measured at
liquid helium temperatures. It was discovered that the inelastic noise
correction is positive for zero-bias conductance close to $1G_0$, but negative
below $0.95G_0$ \cite{Kumar2012}. The transition of the inelastic noise
corrections from positive to negative at $\tau_-$, as predicted by the SLHM,
has not been confirmed experimentally yet.

Interestingly, pioneering first-principles calculations of shot noise characteristics in Ref.~\cite{Avriller2012} could not identify the sign change in the correction to the inelastic shot noise, reported in Ref.~\cite{Kumar2012}. The studied junction structures showed generally transmissions $\tau>0.9$, i.e.\ above the theoretically expected inelastic sign crossover threshold $\tau_+$ \cite{Kumar2012} but below the experimentally reported value of $0.95G_0$ \cite{Kumar2012}. Based on their calculations, the authors speculated that the crossover might occur at a value even lower than $\tau_-$ \cite{Avriller2012}.

Hence, there remain several open points. They regard the confirmation of the sign changes at $\tau_-$ and $\tau_+$ in ab-initio models as well as the experimental verification of the lower threshold at $\tau_-$. On a more quantitative level it is also of relevance to explain the discrepancy between the theoretical sign threshold of $\tau_+$ and the measured one at $\tau\approx0.95$ \cite{Kumar2012}.

In this work we address the aforementioned challenges by investigating inelastic effects in the shot noise of atomic contacts and molecular junctions theoretically and experimentally. Within our newly developed ab-initio approach we detect the noise-sign crossover threshold at a total transmission between $\tau = 0.90$ and $0.95$ for Au atomic contacts. By numerically studying 1,4-benzenedithiol (BDT) connected to gold electrodes, we predict the crossover of the inelastic noise correction in the low conductance range to occur around $\tau\approx0.20$. We propose the Au-BDT-Au single-molecule junctions as a system with a widely tunable conductance in order to observe also this low-transmission sign crossover. Finally, we analyze the influence of vibrational heating on the current-noise properties in Au and Au-BDT-Au junctions theoretically, showing a transition from a linear to a quadratic dependence with increasing voltage, depending on the assumed strength of the coupling of the junctions' vibrational modes to an external reservoir. Our theoretical predictions for Au metallic contacts are supplemented by new experimental data, confirming earlier measurements by Kumar et al.~\cite{Kumar2012}.

This paper is organized as follows. In Sec.~\ref{sec:Methods} we will introduce the theoretical and experimental methods that we use to analyze shot noise in atomic and molecular junctions. We will put a particular emphasis on inelastic shot noise contributions due to EV interactions. In Sec.~\ref{sec:Results} we will present the results and will discuss the two basic systems under study, namely Au atomic junctions and single-molecule contacts made from BDT connected to Au electrodes. We finally summarize our work in Sec.~\ref{sec:Conclusions}.

\section{Methods}\label{sec:Methods}
In this section we present the theoretical and experimental approaches that we apply to study the current and shot noise, including both elastic and inelastic contributions.

\subsection{Theory}
\label{sec:MethodTheory}

\subsubsection{Current and shot noise}

We regard the nanoscale junction as a central device region (C), which is connected to semi-infinite crystalline electrodes to the left (L) and the right (R). The C part consists of the atomic or molecular system and parts of the electrodes at the narrowest constriction. The Hamiltonian of the nanojunction is described by
\begin{equation}
  \hat H = \hat H^{\mathrm{e}} + \hat H^{\mathrm{v}} + \hat H^{\mathrm{ev}},
  \label{eq:Hfull}
\end{equation}
where
\begin{align}
  & \hat H^{\mathrm{e}}=\sum_{ij} \hat d^\dagger H_{ij}^{\mathrm{e}} \hat d_j, \label{eq:He}\\ &
  \hat H^{\mathrm{v}}=\sum_{\alpha}\hbar \omega_{\alpha} \hat b_{\alpha}^\dagger
  \hat b_{\alpha}, \label{eq:Hv}\\ &
  \hat H^{\mathrm{ev}}=\sum_{ij}\sum_{\alpha}\hat
  d^\dagger_i\lambda^{\alpha}_{ij} \hat d_j( \hat b_{\alpha}^\dagger+ \hat b_{\alpha}).  \label{eq:Hev}
\end{align}
Here $H_{ij}^{\mathrm{e}}=\langle i|\hat H^{\mathrm{e}}|j \rangle$ are the
matrix elements of the equilibrium single-electron Hamiltonian $\hat
H^{\mathrm{e}}$ in the nonorthogonal atomic-orbital basis $\{ |i\rangle \}$
with overlap matrix elements $S_{ij}=\langle i | j \rangle$, $\omega_{\alpha}$
is the vibrational frequency of normal mode $\alpha$, and
$\lambda^{\alpha}_{ij}$ are the EV coupling constants
\cite{Viljas2005,Buerkle2013}, which connect two electronic atomic orbitals
with a vibrational mode. $\hat d^\dagger_i$ ($\hat d_i$) and $\hat
b_{\alpha}^\dagger$ ($\hat b_{\alpha}$) are the electron and vibration
creation (annihilation) operators. We determine all the parameters
$H_{ij}^{\mathrm{e}}$, $\omega_{\alpha}$ and $\lambda^{\alpha}_{ij}$ from
parameter-free first-principles calculations in the framework of DFT, as we
explained in detail in a previous study \cite{Buerkle2013}. Further
information on our DFT calculations will be given further below.

To describe charge transport, we use the NEGF technique. In line with previous
approaches \cite{Viljas2005,Frederiksen2007,Haupt2010,Buerkle2013}, we take
vibrations and EV interactions into account only in the C region. The central
quantity, from which all elastic transport properties are extracted, is the
noninteracting retarded Green's function of the C part
\begin{equation}
  \boldsymbol{G}^r(E)=\big[E
    \boldsymbol{S}_{\mathrm{CC}}-\boldsymbol{H}_{\mathrm{CC}}^{\mathrm{e}}-\boldsymbol{\Sigma}_{\mathrm{L}}^{\mathrm{r}}-\boldsymbol{\Sigma}_{\mathrm{R}}^{\mathrm{r}}]^{-1}.\label{eq:Gr}
\end{equation}
In the expression, $\boldsymbol{H}_{\mathrm{CC}}^{\mathrm{e}}$ and
$\boldsymbol{S}_{\mathrm{CC}}$ are the Hamiltonian and overlap matrices of the
C region. The embedding self-energy of the semi-infinite electrode $Z=\mathrm{L},\mathrm{R}$ is given by
\begin{equation}
  \boldsymbol{\Sigma}_{Z}^{\mathrm{r}}(E)=( \boldsymbol{H}_{\mathrm{C}Z}^{\mathrm{e}}-E
    \boldsymbol{S}_{\mathrm{C}Z} ) \boldsymbol{g}^{\mathrm{r}}_{ZZ}(E) (
    \boldsymbol{H}_{Z\mathrm{C}}^{\mathrm{e}}-E \boldsymbol{S}_{Z\mathrm{C}} ),\label{eq:SigmaZ}
\end{equation}
where $\boldsymbol{g}^{\mathrm{r}}_{ZZ}(E)=\big[(E+i\epsilon)\boldsymbol{S}_{ZZ}-\boldsymbol{H}_{ZZ}^{\mathrm{e}} \big]^{-1}$ is the corresponding surface Green's function with the positive infinitesimal broadening $\epsilon$. From Eq.~\eqref{eq:SigmaZ} the linewidth-broadening matrix $\boldsymbol{\Gamma}_{Z}(E)=-2\mathrm{Im}\boldsymbol{\Sigma}_Z^{\mathrm{r}}(E)$ can be obtained. By using the so-called "wide-band limit" (WBL) \cite{Viljas2005,Buerkle2013}, we neglect the energy dependence of the noninteracting Green's functions and related quantities and simply evaluate them at the Fermi energy $E=E_{\text{F}}$.

The EV interaction gives rise to a further self-energy $\boldsymbol{\Sigma}_{\mathrm{ev}}^{\mathrm{r}}=\boldsymbol{\Sigma}_{\mathrm{F}}^{\mathrm{r}}+\boldsymbol{\Sigma}_{\mathrm{H}}^{\mathrm{r}}$ that we split up in Hartree and Fock parts. Then the full retarded Green's function is defined as $\tilde{\boldsymbol{G}}^{\mathrm{r}}(E)=\big[E \boldsymbol{S}_{\mathrm{CC}}-\boldsymbol{H}_{\mathrm{CC}}^{\mathrm{e}}-\boldsymbol{\Sigma}_{\mathrm{L}}^{\mathrm{r}}-\boldsymbol{\Sigma}_{\mathrm{R}}^{\mathrm{r}}-\boldsymbol{\Sigma}_{\mathrm{ev}}^{\mathrm{r}}]^{-1}$. The EV interaction is often very weak. As a result we assume that a perturbative approach to the lowest order in the EV coupling is appropriate for evaluating both the current and shot noise. If we perform the lowest-order expansion with respect to the matrix of EV couplings $\boldsymbol{\lambda}^{\alpha}$ in the C part, we obtain $\tilde{\boldsymbol{G}}^{\mathrm{r}} \approx \boldsymbol{G}^{\mathrm{r}}+\boldsymbol{G}^{\mathrm{r}}\boldsymbol{\Sigma}_{\mathrm{ev}}^{\mathrm{r}}\boldsymbol{G}^{\mathrm{r}}$. In this way the current through the atomic-scale junction can be expressed in terms of the standard elastic contribution and a part that stems from the EV coupling $I=I_{\mathrm{el}}+I_{\mathrm{ev}}$ \cite{Viljas2005,Buerkle2013}. The contribution $I_{\mathrm{ev}}$ is in turn conveniently split into elastic and inelastic corrections according to $I_{\mathrm{ev}}=\delta I_{\mathrm{el}}+I_{\mathrm{inel}}$ \cite{Viljas2005,Buerkle2013}. Analogously the total noise can be expressed as $S=S_{\mathrm{el}}+S_{\mathrm{ev}}$, where $S_{\mathrm{el}}$ is the elastic contribution to the noise, as given by Eq.~\eqref{eq:Sel} and discussed in the introduction, and $S_{\mathrm{ev}}=S_{\mathrm{F }}+S_{\mathrm{H}}$ specifies the respective correction, arising from the Fock (F) and Hartree (H) EV self-energy \cite{Haupt2010}.

To compute inelastic shot-noise signals due to EV scattering in nanoscale junctions, we have implemented the expressions of Haupt et al.\ \cite{Haupt2010}. We have used them to extend our existing code, originally developed to calculate IET spectra \cite{Buerkle2013}. Using the lowest-order expansion in the EV couplings and taking into account only the self-energy corrections due to the Fock diagram, the inelastic correction to the current noise at finite temperature can be written as a summation of the mean-field contribution $S_{\mathrm{F}}^{(\mathrm{mf})}$ and the vertex correction $S_{\mathrm{F}}^{(\mathrm{vc})}$, i.e.\ $S_{\mathrm{F}}= S_{\mathrm{F}}^{({\mathrm{mf}})}+S_{\mathrm{F}}^{({\mathrm{vc}})}$. We use the WBL expressions for finite temperature $T$ and have optimized the numerical evaluations for liquid helium temperatures around $T=4.2$~K. Following Refs.~\cite{Haupt2010,Avriller2012}, we neglect both the correction to noise due to the Hartree diagram and the asymmetric contributions. This is justified as follows. The Hartree part has no effects on the shot noise at the vibrational excitation energies. The asymmetry terms, on the other hand, are much smaller than the symmetric ones at the low temperatures considered here. Since the formulas for $S_{\mathrm{ev}}$ are rather lengthy, we do not reproduce them here, but refer the reader to Ref.~\cite{Haupt2010}.

It is has been reported that the current-driven fluctuations of the phonon occupation lead to a distinct correction to the noise \cite{Urban2010,Novotny2011}. While we ignore such complex effects here, we still consider the influence of vibrational heating on the current noise. To do so, we replace the equilibrium vibrational occupation, given by the Bose distribution $n(E)=[\exp(E/k_{\rm B}T)-1)]^{-1}$, with the nonequilibrium voltage- and temperature-dependent vibrational distribution function
\cite{Viljas2005,Buerkle2013}
\begin{equation}
  N_{\alpha}(E)=\frac{1}{2}\frac{\mathrm{Im}\Pi_{\alpha}^<(E)-n(E)\eta
    E/E_{\alpha}}{\mathrm{Im}\Pi_{\alpha}^{\mathrm{r}}(E)-\eta E/(2E_{\alpha})}.
  \label{eq:Nalpha}
\end{equation}
Here
\begin{equation}
  \Pi_{\alpha}^<(E)=-\frac{i}{2\pi}\int
  dE'\mathrm{Tr}[\boldsymbol{\lambda}^{\alpha}\boldsymbol{G}^<(E')\boldsymbol{\lambda}^{\alpha}\boldsymbol{G}^{>}(E'-E)]
  \label{eq:Pilesser}
\end{equation}
and
\begin{align}
  \begin{aligned}
    \Pi_{\alpha}^{\mathrm{r}}(E)=-\frac{i}{2\pi}\int dE'\mathrm{Tr}[\boldsymbol{\lambda}^{\alpha}\boldsymbol{G}^<(E')\boldsymbol{\lambda}^{\alpha}\boldsymbol{G}^{\mathrm{a}}(E'-E)\\
      +\boldsymbol{\lambda}^{\alpha}\boldsymbol{G}^{\mathrm{r}}(E')\boldsymbol{\lambda}^{\alpha}\boldsymbol{G}^<(E'-E)]
    \label{eq:Piretarded}
  \end{aligned}
\end{align}
are the lesser and the retarded vibrational self-energies, and $\eta$ is a phenomenological parameter, describing the effect of the coupling of the vibrational modes to an external bath, which is provided by the leads. The Green's functions that appear in Eqs.~\eqref{eq:Pilesser} and \eqref{eq:Piretarded} are defined as in Eq.~\eqref{eq:Gr}, $\boldsymbol{G}^{\mathrm{a}}(E)=\boldsymbol{G}^{\mathrm{r}}(E)^\dagger$ and 
\begin{equation}
  \boldsymbol{G}^\lessgtr(E)=\boldsymbol{G}^{\mathrm{r}}[\boldsymbol{\Sigma}^\lessgtr_{\mathrm{L}}(E)+\boldsymbol{\Sigma}^\lessgtr_{\mathrm{R}}(E)]\boldsymbol{G}^{\mathrm{a}}(E)
\end{equation}
with
\begin{align}
  \boldsymbol{\Sigma}^<_Z(E)&=i\boldsymbol{\Gamma}_Z(E)f_Z(E),\\
  \boldsymbol{\Sigma}^>_Z(E)&=i\boldsymbol{\Gamma}_Z(E)[f_Z(E)-1],
\end{align}
where $f_Z(E)=\{\exp[(E-\mu_Z)/k_BT]+1\}^{-1}$ is the Fermi function of lead $Z$. Note that $\eta$ is the only free parameter in our ab-initio model. Motivated by the results in Ref.~\cite{Viljas2005}, we set it to $\eta = 10^{-3}$~eV, unless otherwise noted. In addition we will use in all our transport calculations a temperature of $T=4.2$~K.

\subsubsection{Ab-initio electronic structure}

All the parameters of the Hamiltonian in Eq.~(\ref{eq:Hfull}) are obtained in the framework of DFT, as explained in detail in previous work \cite{Pauly2008,Buerkle2013}. We use DFT as implemented in the TURBOMOLE software package \cite{TURBOMOLE} to optimize geometries, to evaluate the matrix elements of the effective single-particle Hamiltonian $H_{ij}^{\mathrm{e}}$ of the equilibrium structure and to determine vibrational energies $\hbar \omega_\alpha$ and EV couplings $\lambda_{i,j}^\alpha$ in the C part of the nanojunctions. All our calculations are performed with the PBE exchange-correlation functional \cite{Dirac1929,Perdew1996,Perdew1992,Slater1951} and the def-SV(P) basis set \cite{Schaefer1992,Eichkorn1995,Eichkorn1997}, which is of split-valence quality with polarization functions on all non-hydrogen atoms.

\subsection{Experiment}

In order to investigate the influence of EV interaction on the excess noise
$S(V)-S(V=0)$, we basically proceed along the lines of Kumar et
al.\ \cite{Kumar2012}. Differences of our fully functional but simplified
electronic circuitry are explained in \cite{Karimi2016a}. With the help of the
mechanically controllable break-junction technique Au atomic contacts are
formed. Since we are interested in the shot noise of Au single-atom contacts,
all of the measurements are performed on contacts with $G \approx G_0$. At
variance to Kumar et al.\ \cite{Kumar2012} we use thin-film break
junctions. In our thin-film break junctions we obtain single-atom contacts
with conductance values in the range of $0.6 < G/G_0 < 1$ \cite{Scheer2001}.

The shot noise of Au atomic junctions is acquired at $T=4.2$~K, using a
current-amplifier setup \cite{Karimi2016a}. After identifying a stable
contact, the current-voltage characteristics, the differential conductance
$(dI/dV)$, the point-contact spectra $(d^2I/dV^2)/(dI/dV)$ and the shot noise
$S$ are measured. The current of the junction is amplified using a
transimpedance amplifier. Then the noise spectrum between 1 and 100~kHz is
measured with the help of a spectrum analyzer. To determine the shot noise, we
average over a frequency range from 30 to 80~kHz, i.e.\ between $1/f$ noise at
low frequencies and roll-off of the spectra at high frequencies. Finally, the
shot noise is calculated as the excess noise $S(V) - S(0)$ by subtracting the
thermal noise $S(V=0)$ of the junction and the whole setup from the total
noise $S(V)$ at applied bias voltage $V$.

\section{Results and discussion}\label{sec:Results}

In this section we present both theoretical and experimental results for the shot noise of Au atomic junctions. We discuss in particular the transition from negative to positive inelastic shot noise corrections at the high-conductance threshold of the SLHM at $G_0\tau_+$ \cite{Avriller2009,Schmidt2009,Haupt2009}. We then proceed to the purely theoretical results on Au-BDT-Au junctions, which we propose as a system to study also the low-conductance threshold at $G_0\tau_-$. Finally, we discuss the bias-dependent behavior of shot noise for the purely metallic atomic contacts as well as the molecular junctions.

Based on the form of the elastic shot noise in Eq.~\eqref{eq:Sel} and following Ref.~\cite{Kumar2012}, we define the reduced noise $Y(V)=[S(V)-S(0)]/S(0)$ as the difference of the noise at finite bias and that at zero bias, scaled by the zero-bias noise. We represent the voltage dependence at a given temperature by the dimensionless quantity $X(V)=\beta eV \coth(\beta eV/2)/2$. In this way Eq.~\eqref{eq:Sel} is expressed as the simple linear relationship $Y(V)=F[X(V)-1]$.

For Au contacts in the calculation as well as in the experiment we typically
observe piecewise linear dependencies of $Y(X)$ with slopes that change below
and above the threshold for the excitation of a dominant vibrational mode. To
quantify the inelastic correction to shot noise based on the EV coupling, we
therefore determine the slopes $F_1$ and $F_2$ of $Y$ versus $X$ before and
after the kink, respectively, and define the relative change $\delta
F/F=(F_2-F_1)/F_1$ \cite{Kumar2012}.

\subsection{Au atomic contacts}
\subsubsection{Theory}

\begin{figure*}[!t]
  \centering
  \includegraphics[width=0.9\textwidth]{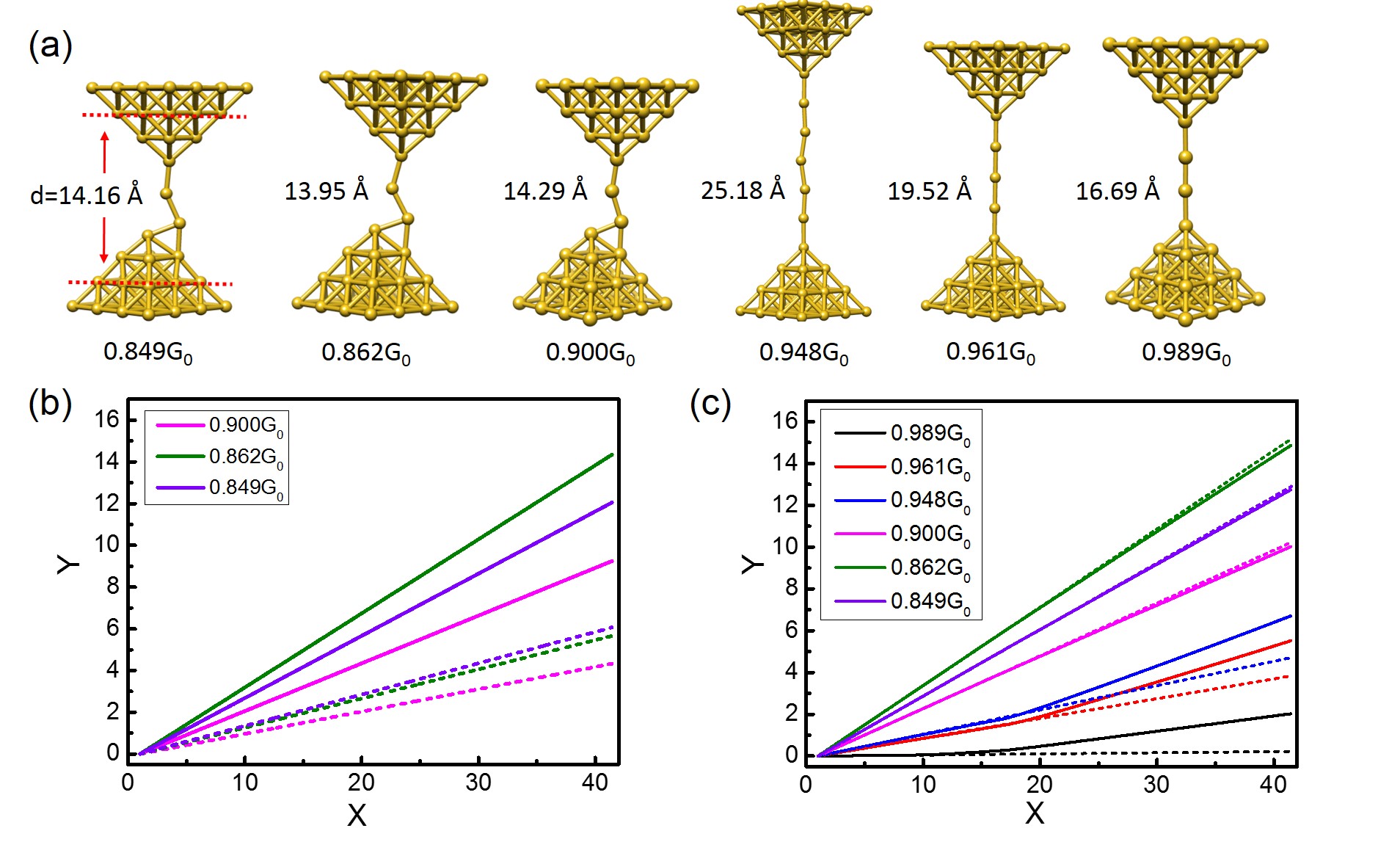}
  \caption{(a) Simulated Au single-atom junctions, consisting of atomic chains of different lengths. The chains bridge the gap between electrodes that are oriented along the $\langle 100 \rangle$ direction. The electrode separation $d$ is defined at the first junction. Together with the elastic conductance $G$ of Eq.~\eqref{eq:G} it is indicated as a label for all geometries. (b) Calculated elastic shot noise in the $Y$-$X$ representation, considering multiple transmission eigenchannels $G = G_0\tau=G_0\sum_{n=1}^N \tau_n$ (solid lines) or just a single transmission eigenchannel $G = G_0\tau=G_0\tau_1$ with the same total transmission as the corresponding multichannel case (dashed lines). (c) Calculated full noise signal (solid lines), containing both elastic and inelastic contributions, and extrapolation of the low-voltage noise characteristics (dashed lines) for the six contacts, shown in panel (a).}
  \label{fig:Au100}
\end{figure*}

\begin{figure*}[!t]
  \centering \includegraphics[width=0.9\textwidth]{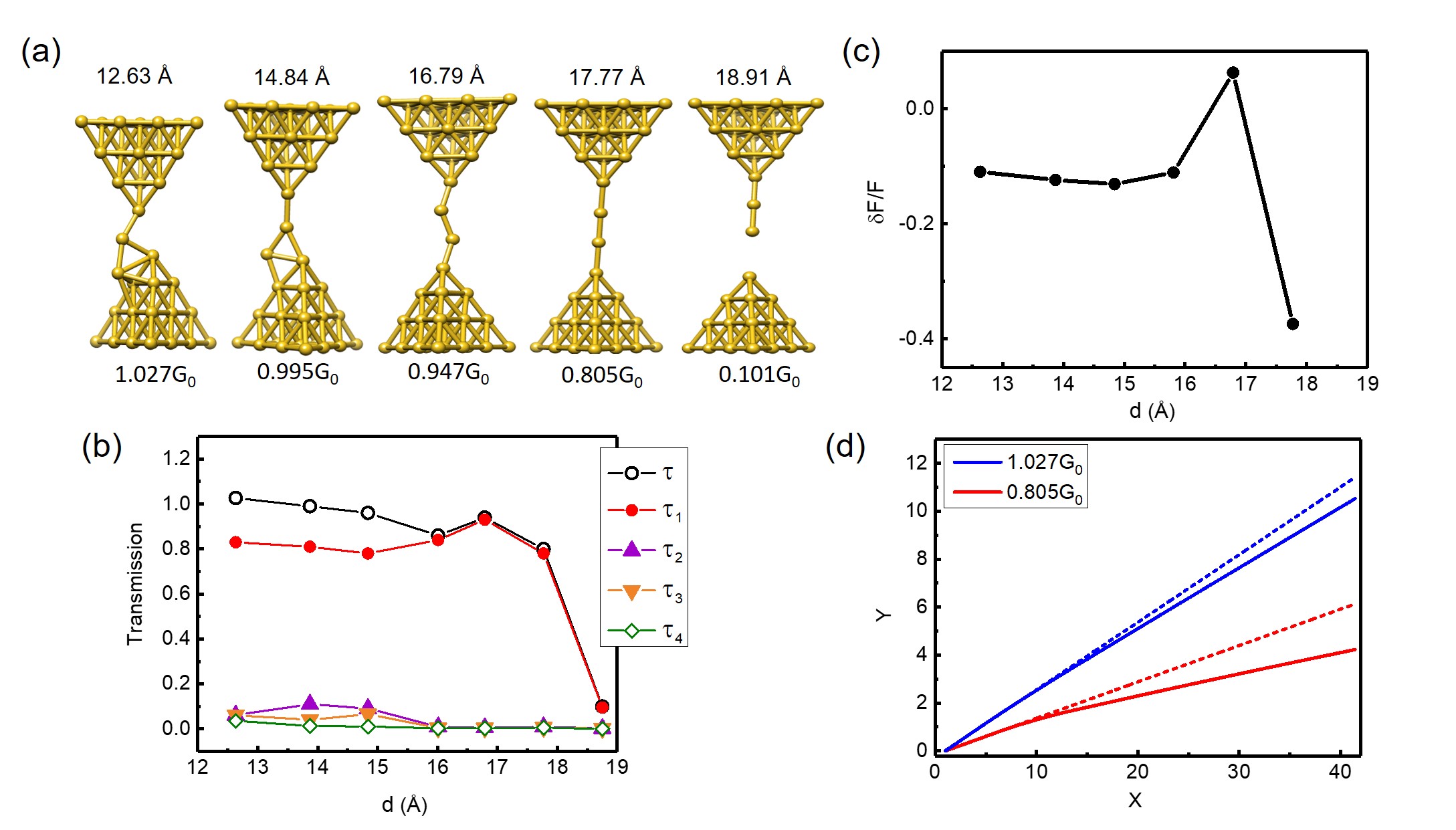}
  \caption{(a) Selected stages of the evolution of an Au wire upon stretching. Electrodes are oriented such that transport proceeds along the $\langle 111 \rangle$ crystallographic direction. Geometries are labeled with the electrode separation $d$ (see Fig.~\ref{fig:Au100} and the elastic conductance $G$ [see Eq.~\eqref{eq:G}]. (b) The total transmission $\tau$ and those of the four largest transmission eigenchannels $\tau_1$ to $\tau_4$ as a function of distance $d$. (c) Relative change in the Fano factor as a function of $d$. (d) $Y$-$X$ representation of the shot noise for Au junctions at elongations $d=12.63$~\AA\ ($G=1.03G_0$) and $d=17.77$~\AA\ ($G=0.81G_0$). Solid lines show the calculated full noise signal, containing both elastic and inelastic contributions, while dashed lines are the extrapolation of the low-voltage noise characteristics.}
  \label{fig:Au111}
\end{figure*}

Gold atomic contacts serve as ideal test systems, for which high-quality
experimental data of inelastic transport is available
\cite{Agrait2002,Kumar2012}. Since transport in atomic-scale junctions is
known to depend crucially on the position of individual atoms
\cite{Agrait2003}, it is important to generate a representative ensemble of
atomic junction geometries for a meaningful comparison with experiment. We
have therefore set up Au atomic junctions under various strain conditions,
exhibiting chains of different lengths at their narrowest cross section. The
atomic chains consist of 1 to 7 atoms, as experimentally reported
\cite{Yanson1998}, and connect two pyramid-shaped Au electrodes in various
configurations. The transport direction in the electrodes coincides either
with the $\langle 100 \rangle$ crystal direction, as shown in
Fig.~\ref{fig:Au100}, or with the $\langle 111 \rangle$ direction, see
Fig.~\ref{fig:Au111}, since transmission electron microscopy studies have
found these crystallographic directions to form in the last stage of the
stretching process \cite{Rodrigues2000}.  We optimize all the atoms in the C
region of each junction, which also represents the "dynamical region", where
atoms can move and vibrations are taken into account. For the first junction
of Fig.~\ref{fig:Au100}(a) the C region consists of all the atoms, located
between the red dotted horizontal lines. In contrast we keep those layers
fixed that we attribute to the L and R electrodes. We measure the electrode
separation $d$ between the first gold layers on each side of the junction that
are kept fixed and that are closest to the C region.

Before we consider the full shot noise signal with inelastic contributions, let us discuss differences in the elastic noise for multichannel and single-channel situations. In Fig.~\ref{fig:Au100}(b) we study three junctions of Fig.~\ref{fig:Au100}(a) with total transmissions of $\tau=0.85$, $0.86$ and $0.90$. Their four highest eigenchannel transmissions $\tau_1$ to $\tau_4$ are specified in Table~\ref{tb:Au100}. We find that most of the systems under study exhibit by a single nearly fully transmitting channel. Others contribute with transmissions within the range of 10\% of the main channel. If we consider all the channel transmissions in Eq.~\eqref{eq:Sel} [solid lines in Fig.~\ref{fig:Au100}(b)], we find a higher noise as compared to a hypothetical case of a single channel with $\tau=\tau_1$ [dashed lines in Fig.~\ref{fig:Au100}(b)]. This is expected, since the elastic noise decreases monotonically for a single channel with $1/2\leq \tau_1 \leq 1$, as $\tau_1$ approaches 1. For the same reason the noise of the contact with $\tau=0.86$ is expected to be below those of the contact with $\tau=0.85$. This ordering is indeed obeyed, if we consider just a single channel. But it is reversed in the multichannel case, since $\tau_1=0.73$ for the particular contact with $\tau=0.86$ is lower than $\tau_1=0.77$ for the contact with $\tau=0.85$. This discussion shows that the channels with small transmission can have a significant influence on the shot noise. In this regard single-channel models may miss some interesting physical effects as compared to multichannel ones.

Our full shot noise results for all of the contacts of Fig.~\ref{fig:Au100}(a) are shown in Fig.~\ref{fig:Au100}(c). By considering different geometrical configurations, we cover a wide range of junction conductances $G$ between $0.85G_0$ and $0.99G_0$. The range of total transmissions is indeed wide enough to observe the transition from a negative to a positive inelastic correction to the noise. Due to the sampling with a limited amount of six geometries, we find the sign crossover to occur between $G=0.90G_0$ and $0.95G_0$. We also see that the inelastic signal does not affect the shot noise of the junctions very much that we studied in Fig.~\ref{fig:Au100}(b). In particular the counterintuitive ordering of the size of their shot noise signal remains intact. The inelastic signal is larger for the better conducting junctions with $G=0.95G_0$, $0.96G_0$ and $0.99G_0$.

\begin{table}[!b]
  \caption{\label{tb:Au100} Conductance and transmission of the
    highest four eigenchannels for the contacts studied in
    Fig.~\ref{fig:Au100}(b).}
  \begin{tabular}{ccccc}
    $G$ ($G_0$) \; & \; $\tau_1$ \; & \; $\tau_2$ \; &
    \; $\tau_3$ \; & \; $\tau_4$ \; \\ \hline
    0.900  & 0.836 & $2.86\times10^{-2}$ & $1.81\times10^{-2}$ & $1.49\times10^{-2}$ \\
    0.862  & 0.742 & $5.76\times10^{-2}$ & $3.65\times10^{-2}$ & $1.84\times10^{-2}$ \\
    0.849  & 0.771 & $3.13\times10^{-2}$ & $2.53\times10^{-2}$ & $1.89\times10^{-2}$ \\ \hline
  \end{tabular}
\end{table}

In simple terms the EV interaction in multichannel junctions can modify the scattering within a channel or lead to scattering between channels. These two effects may be referred to as intra- and interchannel scattering, respectively. They have been discussed in the literature \cite{Buerkle2013,Wheeler2013,Ben-Zvi2013}. The theoretical study of B\"urkle et al.~\cite{Buerkle2013} for the current through an Au chain and the experimental observations of Wheeler et al.~\cite{Wheeler2013} for the noise put forward that the sign and magnitude of change at a given voltage due to EV interaction are determined by the transmission of the particular eigenchannel that happens to be strongly coupled to the relevant local vibrational mode in the C part of the many-channel junction. The analysis of inelastic effects in Pt-benzene-Pt and Pt-CO$_2$-Pt molecular junctions with several transmission channels, on the other hand, suggests that the effect of vibrational excitation on conductance can also involve scattering between channels \cite{Ben-Zvi2013}.

\begin{table}[!b]
  \caption{\label{tb:Au111} Conductance and transmission of the
    highest four eigenchannels for selected contacts studied in
    Fig.~\ref{fig:Au111}.}
  \begin{tabular}{ccccc}
    $G$ ($G_0$) \; & \; $\tau_1$ \; & \; $\tau_2$ \; &
    \; $\tau_3$ \; & \; $\tau_4$ \; \\ \hline
    1.027  & 0.835 & $6.75\times10^{-2}$ & $6.15\times10^{-2}$ & $3.62\times10^{-2}$ \\
    {0.947} & 0.933 & $5.51\times10^{-3}$ & $4.25\times10^{-3}$ & $3.69\times10^{-3}$ \\
    0.805  & 0.781 & $9.94\times10^{-3}$ & $7.08\times10^{-3}$ & $6.36\times10^{-3}$ \\ \hline
  \end{tabular}
\end{table}

Atomic-scale configurational changes can have considerable effects on the noise, because of the differing numbers of transmission eigenchannels that might be active in each case, the changing symmetry of vibrational modes, and the size of corresponding EV couplings. We study this aspect in Fig.~\ref{fig:Au111} by stretching a junction with electrodes, oriented along the $\langle 111 \rangle$ direction. At low interelectrode separations $d$ between 12.63 and 14.84~\AA, we obtain contacts with as many as four significant transmission eigenchannels, while a single prevalent channel emerges when we continue the stretching. We observe that the sign of the inelastic noise correction in Fig.~\ref{fig:Au111}(c) can be well understood by the behavior of the dominant transmission channel $\tau_1$, i.e.\ intrachannel EV coupling. When $\tau_1$ is below the value of around $\tau_+$, the value of $\delta F/F$ is negative, but it changes to positive, when $\tau_1=0.93$ at $d=16.79$~\AA\ before contact rupture. We specify the transmission values of the highest four channels of the contacts at $d=12.63$, $16.79$ and $17.77$~\AA\ in Table~\ref{tb:Au111} and show the $Y$-$X$ representations of the shot noise for $d=12.63$ and $17.77$~\AA\ in Fig.~\ref{fig:Au111}(d).

\begin{figure*}[!t]
  \centering
  \includegraphics[width=1.0\textwidth]{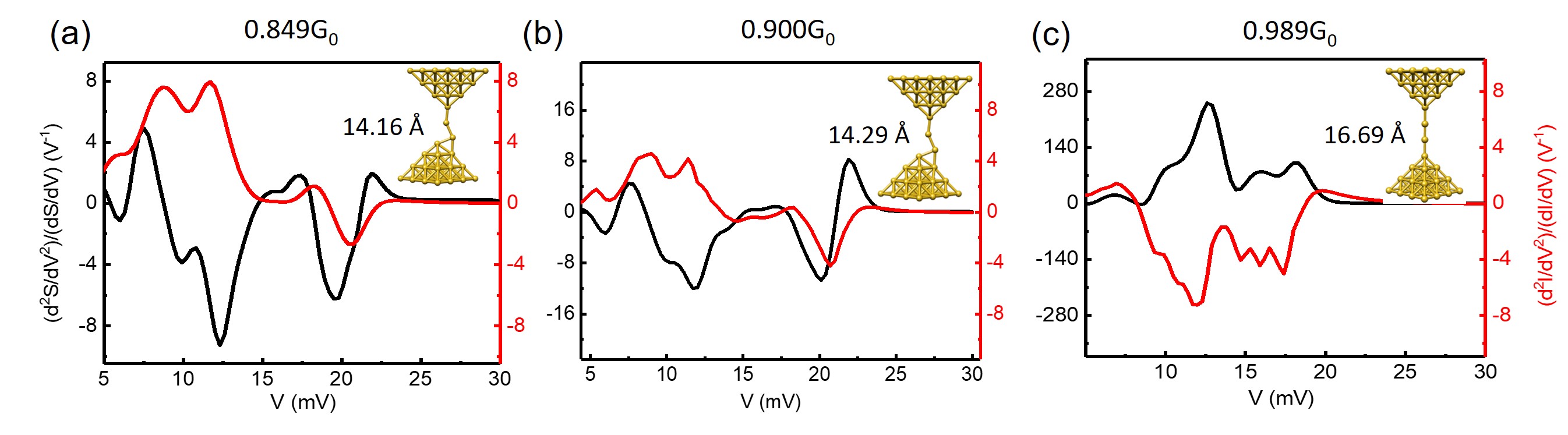}
  \caption{Calculated second derivative of noise and current with respect to voltage for gold junctions with conductance values of (a) $0.85G_0$, (b) $0.90G_0$ and (c) $0.99G_0$. The channel transmissions $\tau_1$ to $\tau_4$ for the contacts in panels (a) and (b) are as specified in Table~\ref{tb:Au100}, while they are $\tau_1=0.967$, $\tau_2=2.38\times 10^{-2}$, $\tau_3=1.67\times 10^{-3}$ and $\tau_4=1.21\times 10^{-3}$ for panel (c). Junction geometries and corresponding electrode separations $d$ are displayed as insets.}
  \label{fig:Au-d2SdV2}
\end{figure*}

As discussed in the introduction, for a symmetric single-channel junction in the SLHM the EV interaction leads to a step up of the conductance for $\tau<1/2$ and a step down for $\tau>1/2$, as the energy $eV$ supplied by the external voltage grows larger than the vibrational energy. In the second derivative of the current with respect to voltage the features appear as a peak or dip, respectively. Along the same line dips are expected at the vibrational activation threshold in the SLHM for the second derivative of the noise $S$ with respect to voltage in the interval $\tau_- < \tau < \tau_+$ and peaks in the other regions $\tau<\tau_-$ or $\tau>\tau_+$. In Fig.~\ref{fig:Au-d2SdV2} we examine these relations for three Au atomic contacts with conductance values of $0.85G_0$, $0.90G_0$ and $0.99G_0$ by plotting both $d^2I/dV^2$ and $d^2S/dV^2$, each normalized by the corresponding first derivative. 

In contrast to the observations in Ref.~\cite{Kumar2012} peaks as well as dips are visible in the IET spectra of Fig.~\ref{fig:Au-d2SdV2}. The high-energy part of the spectra is easier to understand than that at low energies. Around 17 to 22~meV, corresponding to the energy range where longitudinal vibrational modes of Au chains occur, we consistently find dips. Despite $\tau\approx 1$, the IET signals at low voltage between 5 and 15~mV show a positive sign in Fig.~\ref{fig:Au-d2SdV2}(a) and \ref{fig:Au-d2SdV2}(b). In the light of the SLHM this might signal vibrational coupling to eigenchannels with low transmission. Further below, we will see however that this interpretation is incompatible with the behavior of the $d^2S/dV^2$ spectrum. Importantly, as compared to the rather stretched-out geometry in Fig.~\ref{fig:Au-d2SdV2}(c), vibrational modes exhibit no clear symmetry in the bent-chain configurations of Fig.~\ref{fig:Au-d2SdV2}(a) and \ref{fig:Au-d2SdV2}(b), which may lead to complex EV interaction effects. 

Also the inelastic noise contributions yield a complicated
picture. Irrespective of the conductance studied in Fig.~\ref{fig:Au-d2SdV2},
$d^2S/dV^2$ features a pronounced voltage dependence. Overall it shows the
sign change expected for intrachannel scattering in the SLHM, when considering
the size of $\tau_1$. As visible from the data in Table~\ref{tb:Au100},
$\tau_1<\tau_+$ for the geometries in Fig.~\ref{fig:Au-d2SdV2}(a) and
\ref{fig:Au-d2SdV2}(b), and $d^2S/dV^2$ is indeed mainly negative there. For
Fig.~\ref{fig:Au-d2SdV2}(c) with $\tau_1>\tau_+$ in contrast it is mostly
positive. For that latter contact with $G = 0.99G_0$ and $\tau_1\approx 0.97$,
the signals in $d^2I/dV^2$ and $d^2S/dV^2$ have opposite sign, consistent by
the SLHM for this nearly perfect transmission. $d^2S/dV^2$ spectra in
Fig.~\ref{fig:Au-d2SdV2}(a) and \ref{fig:Au-d2SdV2}(b) show both positive and
negative values, which may effectively reduce the integrated signal size of
inelastic shot noise corrections. Negative values of $d^2S/dV^2$ between 5 to
15~mV in Fig.~\ref{fig:Au-d2SdV2}(a) and \ref{fig:Au-d2SdV2}(b) are
inconsistent with vibrational coupling to eigenchannels $n\geq2$ with
$\tau_n<\tau_-$, and the role of interchannel mixing needs further
exploration. Finally, let us focus on the region around 20~meV, where all of
the IET spectra show the expected dip. When going from
Fig.~\ref{fig:Au-d2SdV2}(a) to \ref{fig:Au-d2SdV2}(c), we reveal a transition
from correlation to anticorrelation between $d^2I/dV^2$ and $d^2S/dV^2$ as
$\tau_1$ increases above $\tau_+$ in agreement with the SLHM. Indeed
Fig.~\ref{fig:Au-d2SdV2}(b) represents a transitional stage, where
$\tau_1=0.84$ is very close to $\tau_+$, and $d^2S/dV^2$ displays a dip-peak
feature with a corresponding sign change at the peak in the IET spectrum.

\subsubsection{Experiment}

\begin{figure*}[!tb]
  \centering
  \includegraphics[width=1.0\textwidth]{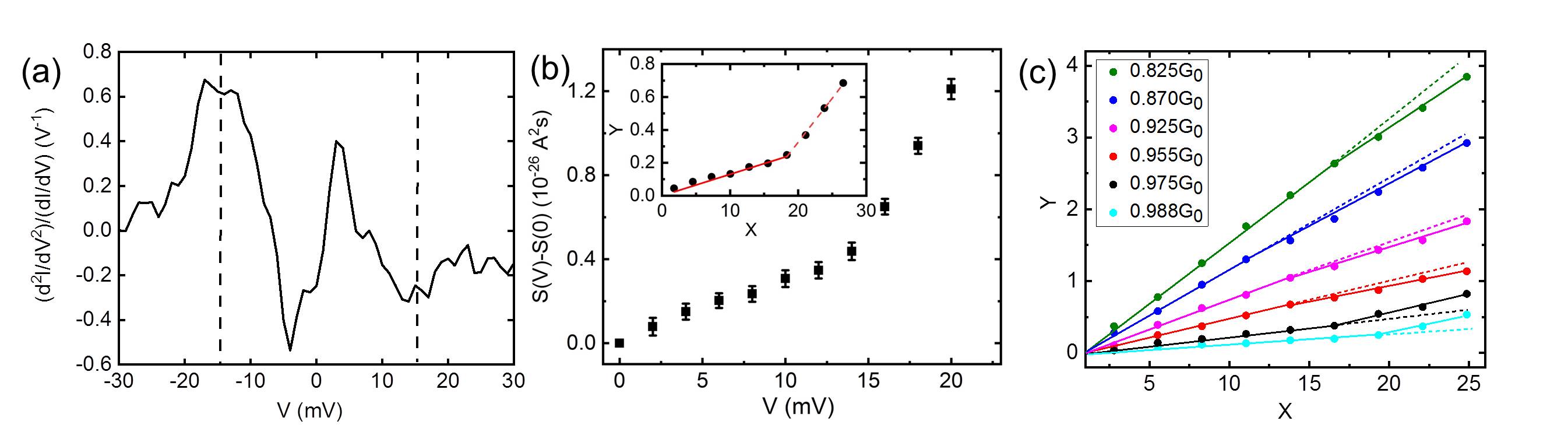}
  \caption{(a) Experimental point-contact spectrum for an Au atomic contact
    with conductance $G=0.987G_0$. The vertical black dashed lines indicate
    the symmetric position of the most prominent vibrational mode at positive
    and negative voltage. (b) The measured current noise for this contact with
    statistical error bars as a function of voltage $V$. The inset shows the
    noise converted to $Y$-$X$ representation. The red solid line is a fit to
    the data points from $X=0$ up to around $20$ (corresponding to voltages
    below 15 mV), from which a Fano factor $F_1=0.013$ is deduced, signaling
    that the contact is realized predominantly by a single eigenchannel with
    $\tau_1 = 0.987$. The linear fit to data points at $X>20$
    (i.e.\ voltages above 15 mV), as indicated by the red dashed line, leads
    to a Fano factor $F_2=0.053$. (c) $Y$-$X$ representation of the shot noise
    for six single-channel contacts, as revealed by the analysis of $F_1$,
    with different conductance values ranging from $0.825G_0$ to
      $0.987G_0$. The solid lines represent the full noise signal, while the
    dashed lines are linear fits to the noise at low $X$.}
  \label{fig:d2IdV2-SV-XY}
\end{figure*}

Fig.~\ref{fig:d2IdV2-SV-XY}(a) shows the point contact spectrum of an experimentally realized Au atomic contact with a linear conductance of $0.987G_0$. The prominent dip at $15\pm1$~mV and corresponding antisymmetric peak at $-15\pm1$~mV indicate an electron scattering process with a vibrational mode. Depending on the atomic configuration, we observe vibrational energies in the range of 10 to 20~meV, which have the tendency to decrease with stretching due to bond softening \cite{Agrait2002}. The peak at around 5~mV is usually attributed to a zero-bias anomaly \cite{Agrait2002,Djukic2006}. But it might also indicate a soft phonon mode \cite{Boehler2007,Boehler2009,Weber2017} and resembles the low-energy features observed in the simulations for the contacts shown in Fig.~\ref{fig:Au-d2SdV2}(a) and \ref{fig:Au-d2SdV2}(b).

Fig.~\ref{fig:d2IdV2-SV-XY}(b) illustrates the measured current noise
$S(V)-S(0)$ of the same contact as a function of bias voltage $V$. The inset
displays the corresponding conversion to the reduced quantities $Y$ and
$X$. The red solid line indicates a linear fit up to the voltage, at which
significant inelastic excitations set in. The data can be well described by
assuming a single channel with a transmission probability of
$\tau=\tau_1=0.987\pm0.002$ and yields the Fano factor $F_1 = 0.013 \pm
0.002$. The dashed line shows the linear fit for voltages above the kink,
yielding the modified "Fano factor" $F_2 = 0.053\pm 0.002$. From these two
values we compute the relative change in the Fano factor $\delta F/F = 3.08
\pm 0.65$, as discussed above.

Finally, Fig.~\ref{fig:d2IdV2-SV-XY}(c) represents the measured shot noise in the $Y$-$X$ reduced units for six contacts with conductances from $0.825G_0$ to $0.987G_0$. The determination of the dominant vibrational mode energy for some of the contacts is difficult due to conductance fluctuations or the mentioned appearance of multiple features in the IET spectra \cite{Boehler2007}. We therefore consistently choose the first minimum in the IET spectra at positive voltage as a signature of a vibrational mode and use this voltage for the location of the kink in the $Y$-$X$ representation of the noise. Solid lines represent the full signal and consist of the two piecewise linear fits to the noise, as described in the previous paragraph.

\subsubsection{Comparison between theory and experiment}

\begin{figure}[tb!]
  \centering
  \includegraphics[width=0.9\columnwidth]{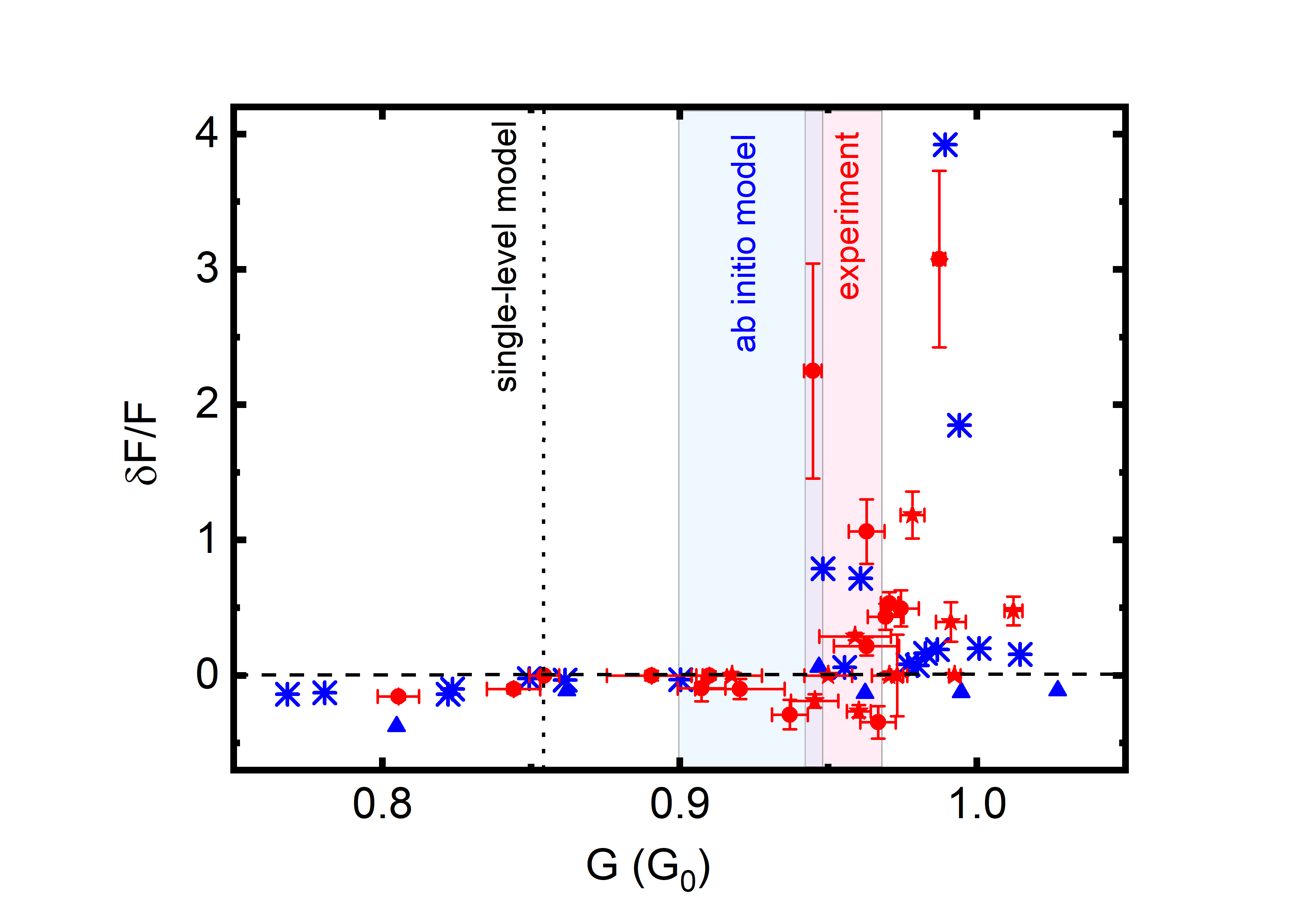}
  \caption{Inelastic shot noise correction as quantified by $\delta F/F$ from experiment and theory. Experimental data for 15 single-channel contacts and 12 double-channel contacts is shown by red cycles and asterisks, respectively, where the errors denote the statistical error resulting from the fitting procedure. The related theoretical results of our ab-initio model are shown in blue. Here, 19 blue stars represent different junctions with leads oriented along the $\langle 100 \rangle$ direction and 6 blue triangles those with leads along $\langle 111 \rangle$. The vertical line as well as the shaded areas indicate the  approximate values and ranges for the transition from negative to positive inelastic shot noise corrections. Thus, the black-dotted line refers to the threshold $\tau_+$ of the SLHM \cite{Avriller2009,Schmidt2009}, the blue-shaded area indicates the lower bound of the threshold found in our ab-initio calculations, and the red-shaded area marks the transition region deduced form our experimental data.}
  \label{fig:dFF}
\end{figure}

We now compare in detail the results for inelastic shot noise corrections found in theory and in experiment. Fig.~\ref{fig:dFF} shows $\delta F/F$ for 27 measured contacts with conductance values between $0.8G_0$ and $1.02G_0$. By comparison between actual $F_1$ values and the expectation for the single-channel case, i.e. $F_1 = 1-\tau_1$, we classify the contacts into single-channel and multichannel ones. We assign a single channel, if the contribution of the additional channels to $F_1$ cannot be revealed within our experimental resolution. We cannot assign a sharp numerical criterion here, because this procedure depends on the precision, with which $G$ can be measured. The latter is limited by conductance fluctuations or soft phonons, as mentioned above. Taking these considerations into account, the lower limit for transmission contributions of additional channels is on the order of 0.003 to 0.01. Single-channel contacts and multichannel ones are marked with different symbols in Fig.~\ref{fig:dFF}. In the same way we plot data from overall 25 theoretically computed contact geometries. Covering the conductance range of $0.76G_0$ to $1.03G_0$, they split into 19 contacts with leads oriented along the $\langle 100 \rangle$ direction and 6 with leads along $\langle 111 \rangle$, which we distinguish by different symbols (see also the corresponding results in Figs.~\ref{fig:Au100} and \ref{fig:Au111}). In the experimental work by Kumar et al.~\cite{Kumar2012} the transition from negative to positive $\delta F/F$ was found around $0.95G_0$. The vertical dotted black line in Fig.~\ref{fig:dFF} represents the value $\tau_+ G_0$ from the symmetrically coupled SLHM of Refs.~\cite{Avriller2009,Schmidt2009}. In our experimental data we find positive as well as negative $\delta F/F$ in a range between $0.94G_0$ and $0.97G_0$, shaded in red in Fig.~\ref{fig:dFF}, while the bluish area is the prediction of our multichannel and multivibration ab-initio modeling.

Due to the finite amount of calculated junction geometries, our ab-initio model locates the sign crossover in the range between $0.90G_0$ and $0.94G_0$. Interestingly, our simulations confirm that the threshold is not sharp and provide an explanation for the data points at $G>0.96G_0$ with negative $\delta F/F$. As discussed in the context of Figs.~\ref{fig:Au111} and \ref{fig:Au-d2SdV2}, they may arise from multichannel junctions. Also in the experimental data we find a trend of negative $\delta F/F$ values for contacts which we identified as multichannel cases, in agreement with the theoretical findings.

Unlike a previous ab-initio study \cite{Avriller2012} we are thus able to observe the high-transmission sign change in the inelastic shot noise correction, which was detected experimentally \cite{Kumar2012} and which we confirm here through independent measurements. By considering the electronic multilevel structure and many vibrational modes, the transmission at the sign change is increased from $\tau_+ G_0$ for the SLHM to a value between $G=0.90G_0$ and $G=0.94G_0$, quite compatible with the experiments. At the same time the transition is seen to be washed out by the electronic multilevel structure, since high-conductance junctions can occur with a relatively low transmissive first eigenchannel, which may cause a negative $\delta F/F$ even for $G>0.96G_0$. This result is apparent only, since we have explored a large set of junction configurations both in experiment and theory. Our theoretical analysis of $d^2S/dV^2$ in Fig.~\ref{fig:Au-d2SdV2} further demonstrates that positive as well as negative inelastic noise contributions with varying weight may arise in a junction, partially averaging out the integrated inelastic signature in $\delta F/F$.

Remaining discrepancies between theory and experiment, for instance with respect to the precise position and width of the crossover area, may be attributed to the limited amount of junctions analyzed. The theory might be further improved by going beyond the WBL approximation, in which the energy dependencies of Green's functions and related quantities are neglected. Furthermore we have concentrated on the symmetric terms of the inelastic noise corrections, as discussed in subsection~\ref{sec:MethodTheory}. From the experimental side, undetected additional noise contributions \cite{Lumbroso2018} or changes of the contacts during the time-consuming noise measurements could affect the determination of the transmissions, and conductance fluctuations superimposed on the vibrationally induced nonlinearities in the point contact spectra might limit the precision of the conductance determination.

\subsection{Au-benzenedithiol-Au contacts}

\begin{figure*}[!tb]
  \centering
  \includegraphics[width=0.8\textwidth]{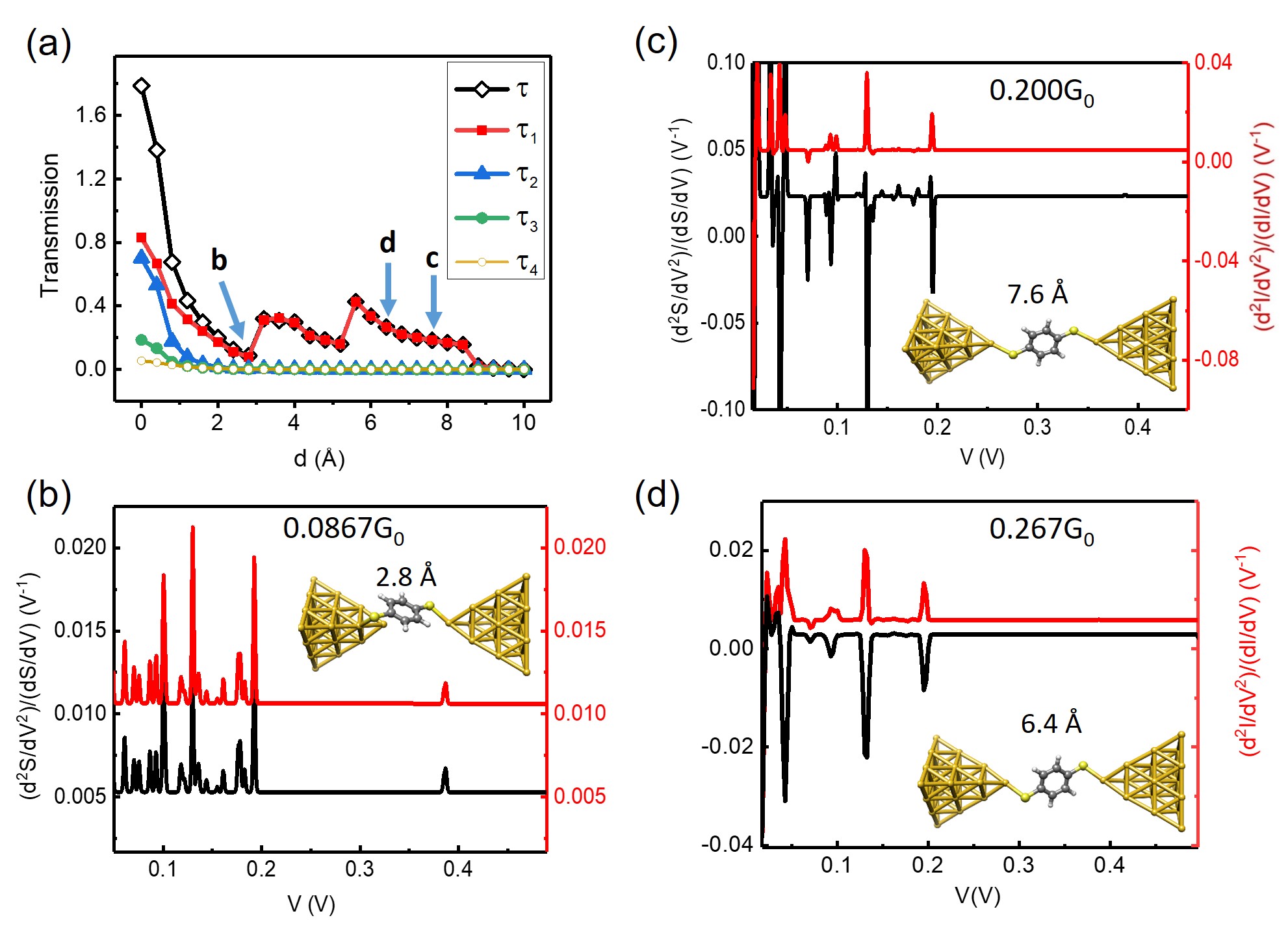}
  \caption{(a) Total transmission and those of the four most transmissive eigenchannels for an Au-BDT-Au junction, as obtained during its stretching. Second derivative of shot noise and current with respect to voltage for Au-BDT-Au junctions with (b) $G=0.087G_0$, (c) $G=0.20G_0$ and (d) $G=0.27G_0$. The selected junctions are shown as insets. They are labeled with the corresponding electrode separation $d$, which is also indicated by arrows in panel (a).}
  \label{fig:BDT-d2SdV2}
\end{figure*}

So far the effect of EV interactions on the noise characteristics of molecular contacts is scarcely studied. The majority of shot noise measurements for single-molecule junctions was carried out at low bias voltages, and they use $S$ to extract information on elastic transmission coefficients, as mentioned above \cite{Djukic2006,Kiguchi2008,Ben-Zvi2013,Karimi2016a}. Only very recently the inelastic contributions to shot noise have been addressed experimentally in these kind of systems \cite{Tewari2019}. However, Ref.~\cite{Tewari2019} concentrated on highly transmissive contacts with a conductance close to $1G_0$. BDT contacted by Au electrodes has been demonstrated to be a system with widely tunable conductance values from $10^{-3}G_0$ to more than $0.5G_0$ \cite{Kim2011a}. Beside inelastic current contributions \cite{Kim2011a} some of the authors reported measurements of the elastic noise in this system \cite{Karimi2016a}, covering a similar conductance range between $10^{-2}G_0$ and $0.24G_0$. The adjustment of the conductance by mechanical control should allow experimental access to the transition from positive to negative inelastic shot noise corrections in the low-conductance regime near $\tau_- G_0$, as we will show theoretically in the following.

We use here the geometries that we have determined in Ref.~\cite{Karimi2016a} during the stretching of a Au-BDT-Au junction to evaluate the inelastic noise. Similar to the gold junctions we analyze in Fig.~\ref{fig:BDT-d2SdV2} the appearance of peaks and dips in the second voltage derivative of current and noise. For this purpose we select three different junction configurations with $G=0.087G_0$, $0.20G_0$ and $0.27G_0$, whose transmission coefficients $\tau_1$ to $\tau_4$ are specified in Table~\ref{tb:BDT}. The progression of the total transmission and those of the largest four eigenchannels is shown in Fig.~\ref{fig:BDT-d2SdV2}(a) as the separation between the electrodes increases. In this case the distance $d$ specifies the displacement of the electrodes with respect to the starting geometry. Since the transmission $\tau$ of the selected geometries remains below $1/2$, we expect that the IET spectra show mainly peaks, while $d^2S/dV^2$ should exhibit a transition from peaks to dips as the conductance increases. This behavior is exactly seen, when going from the low conductance of $0.087G_0$ in Fig.~\ref{fig:BDT-d2SdV2}(b) via the intermediate case at $0.20G_0$ with peaks and dips in $d^2S/dV^2$ in Fig.~\ref{fig:BDT-d2SdV2}(c) to $0.27G_0$ in Fig.~\ref{fig:BDT-d2SdV2}(d). More generally, Fig.~\ref{fig:BDT-d2SdV2} shows that spikes in $d^2S/dV^2$ and $d^2I/dV^2$ correlate in an excellent manner. In addition, we note that both $d^2S/dV^2$ and $d^2I/dV^2$ show a certain offset from zero at finite voltages in Fig.~\ref{fig:BDT-d2SdV2}(a) to \ref{fig:BDT-d2SdV2}(d). It stems from a quadratic background due to phonon heating \cite{Viljas2005,Buerkle2013}, as will be discussed further in subsection \ref{subsec:Phonon-heating}. 

Our results for the different Au-BDT-Au junction geometries show that the change in the noise from positive to negative inelastic corrections in the low-conductance regime occurs slightly above the value $\tau_- G_0$, predicted by the SLHM. As $d^2S/dV^2$ shows both positive and negative values for $0.2G_0$, we attribute the increased threshold conductance to the complex interplay between multiple electronic and vibrational levels coupled via the EV interaction, in analogy to the results for pure Au contacts.

\begin{table}[!bt]
	\caption{\label{tb:BDT} Conductance and transmission of the	highest four eigenchannels for the contacts studied in		Fig.~\ref{fig:BDT-d2SdV2}(b) to \ref{fig:BDT-d2SdV2}(d).}
	\begin{tabular}{ccccc}
		$G$ ($G_0$) \; & \; $\tau_1$ \; & \; $\tau_2$ \; &
		\; $\tau_3$ \; & \; $\tau_4$ \; \\ \hline
		$8.67\times 10^{-2}$  & $7.91\times10^{-2} $ & $5.38\times10^{-3}$ & $1.47\times10^{-3}$ & $3.72\times10^{-4}$ \\
		$0.200$ & $0.200$ & $2.29\times10^{-4}$ & $4.64\times10^{-6}$ & $3.41\times10^{-7}$ \\
		$0.267$ & $0.267$ & $2.72\times10^{-4}$ & $2.55\times10^{-5}$ & $1.48\times10^{-6}$ \\ \hline
	\end{tabular}
\end{table}

\subsection{Heating at high bias voltages}\label{subsec:Phonon-heating}

\begin{figure}[!tb]
  \centering \includegraphics[width=0.8\columnwidth]{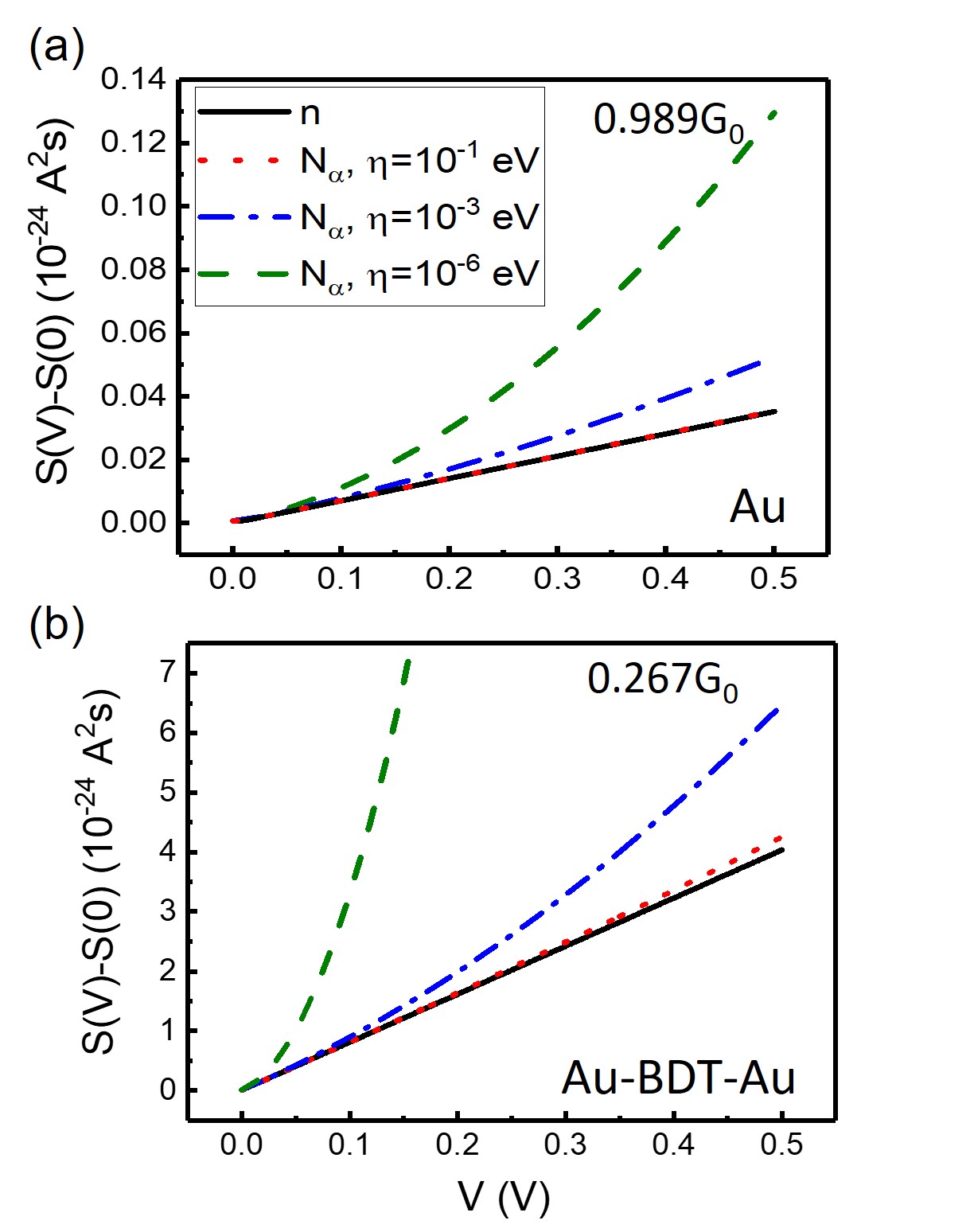}
  \caption{Shot noise of (a) an Au junction with conductance $G=0.99G_0$ and
    (b) an Au-BDT-Au junction with $G=0.27G_0$. As indicated by the legend in panel (a), the equilibrium vibrational distribution $n$ at our default temperature of $T=4.2$~K is compared to effective vibrational occupations $N_{\alpha}(E)$ [see Eq.~\eqref{eq:Nalpha}] that arise from different coupling strengths $\eta$ of vibrational modes in the center of the junction to an external bath.}
  \label{fig:Au-BDT-highV}
\end{figure}

At high bias voltages a nonequilibrium phonon population will be excited in atomic and molecular junctions. Theoretical calculations predict the noise in this regime to grow as $V^\alpha$ with $\alpha\geq2$ due to the coupling between electrons and thermally nonequilibrated vibrational modes \cite{Haupt2010,Urban2010,Novotny2011,Novotny:2015jy}.

As we presented before, we include the effects of vibrational heating on inelastic noise by considering the nonequilibrium vibrational occupation in Eq.~\eqref{eq:Nalpha}. The vibrational broadening $\eta$ describes the coupling of the vibrational modes to an external reservoir, as provided by the electrodes, and the related damping. It is the only free parameter in our theoretical model. Based on the favorable comparison of theoretical IET spectra \cite{Viljas2005} to high-quality experimental data for Au atomic-chain junctions \cite{Agrait2002}, we believe that our default parameter of $\eta=10^{-3}$~eV is a realistic value. Nonetheless, in Fig.~\ref{fig:Au-BDT-highV} we show for Au and Au-BDT-Au junctions with conductances of $0.99G_0$ and $0.27G_0$, respectively, how the shot noise varies with $\eta$ at high bias. For large enough $\eta$, Eq.~\eqref{eq:Nalpha} reduces to the equilibrium Bose distribution, and the shot noise increases linearly with voltage. If $\eta$ is reduced sufficiently, however, we find a superlinear noise curve that can be fitted by a linear plus a quadratic term. 

The fact that we can explain the experimental data for Au atomic junctions in Fig.~\ref{fig:d2IdV2-SV-XY}(c) by piecewise linear fits means that there is no significant influence of thermally nonequilibrated vibrational modes on $S$ at the low biases measured. A quadratic increase in shot noise, reminiscent of the theoretical predictions with low $\eta$, has however been observed experimentally at room temperature for a junction conductance of $3G_0$ and for applied voltages up to around $0.35$~V \cite{Chen2014}. Subsequent studies at low temperatures between 4.2 and 100~K rather emphasize the fact that EV interactions and heating of phonons in atomic contacts are weak \cite{Chen2016}. Our results confirm these experimental studies.

\section{Conclusions}\label{sec:Conclusions}

In summary, based on the NEGF technique and using a Hamiltonian parameterized from DFT, we have studied inelastic effects due to EV coupling on the current noise in systems with multiple electronic levels and vibrational modes. Sign crossover thresholds for inelastic noise are observed at conductances of $G\approx0.2G_0$ and $G\approx0.90G_0$-$0.95G_0$ for Au-BDT-Au single-molecule and pure Au single-atom contacts, respectively. As compared to the SLHM that predicts values of $G=\tau_-G_0\approx0.15G_0$ and $G=\tau_+G_0\approx0.85G_0$, respectively, this increase can be understood by the presence of several partially open transmission eigenchannels that contribute to the total transmission in addition to a dominant one and couple differently to various vibrations. In other words since the inelastic signals are mainly determined by the highest conduction channel and since the transmission of the dominant channel is always lower than the total transmission $\tau_1<\tau$, the apparent inelastic sign thresholds are shifted towards higher $G$ than expected from the simplified single-level toy model.

We have also reported shot noise measurements for Au contacts, using the
mechanically controllable break-junction technique and applying a custom-made,
versatile setup with simplified measurement electronics
\cite{Karimi2016a}. The measurements show nonlinearities in the shot noise
power for bias voltages around corresponding characteristic vibrational mode
energies. The observed crossover from positive to negative sign of inelastic
shot noise corrections, as quantified by the relative Fano factor $\delta
F/F$, occurs in a range between $0.93G_0$ and $0.97G_0$. Our findings confirm
previous experimental results \cite{Kumar2012}, in which the crossover was
located at $0.95G_0$. We conclude that the deviation between the analytically
predicted crossover at $\tau_+$ for a single channel and the experimental
observation for Au contacts may be explained by the occurrence of multichannel
contacts in the experiment. Multiple transmissive eigenchannels also provide a
natural explanation for why the transition is not sharp. Indeed we find
negative values of $\delta F/F$ for $G>0.96G_0$ both in experiment and theory,
which can be assigned to contacts with increased transmission values $\tau_n$
for channels $n\geq2$.

Finally, we have theoretically explored the effects of vibrational heating on the current noise properties as a function of the coupling of vibrations to an external reservoir in the electrodes. For low enough coupling we find a quadratic increase of the noise as a function of voltage at large bias. With increasing coupling the bias-dependent noise becomes linear as the nonequilibrium distribution approaches the equilibrium one. This behavior is similar both for Au single-atom and Au-BDT-Au single-molecule junctions.

The inelastic sign crossover in the noise at low conductance values $G_0\tau_-$ could not be measured yet and thus remains to be verified. The challenging experiments employing high bias voltages are expected to reveal important insights into charge transport through nanosystems beyond elastic theories.

\section{Acknowledgments}

We thank Juan Carlos Cuevas for stimulating discussions and Federica Haupt for scientific exchange. All authors acknowledge financial support through the Collaborative Research Center (SFB) 767 of the German Research Foundation (DFG). In addition F.P.\ thanks the Carl Zeiss Foundation for funding. An important part of the numerical modeling was carried out on the computational resources of the bwHPC program, namely the bwUniCluster and the JUSTUS HPC facility.


%

\end{document}